\documentclass[oldversion]{aa}

\usepackage{graphicx}
\usepackage{lscape}

\usepackage{times}
\usepackage{epsfig}
\usepackage{epsf}
\usepackage{rotating}

\begin{document}



\title{GALEX observations of  quasar variability in the ultraviolet}


\author{
Barry Y. Welsh\inst{1,3}, Jonathan M. Wheatley\inst{1,3} and
James D. Neill \inst{2}}

\institute{Space Sciences Laboratory, University of California, 7 Gauss Way, Berkeley, CA 94720 \and
California Institute of Technology, MC 405-47, 1200 East California Boulevard, Pasadena, CA 91125
\and Eureka Scientific Inc., 2452 Delmer Street, Oakland, CA 94602}


\titlerunning{Quasar UV variability}
\authorrunning{Welsh, Wheatley $\&$ Neil}


\abstract
{}
{Using archival observations
recorded over a 5+ year timeframe with the NASA
Galaxy Evolution Explorer ($\it GALEX$) satellite, we present a study of the ultraviolet (UV) variability of 4360 quasars of
redshifts up to $\it z$ = 2.5
that have optical counterparts in the
Sloan Digital Sky Survey DR5 spectroscopic catalog of Schneider et al. (2007). The observed changes in
both the far UV (FUV: 1350 - 1785\AA) and near UV (NUV: 1770 - 2830\AA) AB magnitudes as a function of time
may help differentiate between models of the emission mechanisms thought to operate in
these active galaxies.}
{A list of NUV and FUV  variable quasars was derived from the UV light-curves of sources with 5 or more observational visits by $\it GALEX$
that spanned a time-frame $>$ 3 months.
By measuring the error in the derived mean UV magnitude from the series of $\it GALEX$ observations for each source, quasars whose
UV variability was greater than the 3-$\sigma$ variance from the mean observed value were deemed
to be (intrinsically) UV variable. This conservative selection criterion (which was applied to both FUV and NUV observations)
resulted in identifying 550 NUV and 371 FUV quasars as being
statistically significant UV variable objects.}


\maketitle


\section{Introduction}
The flux emitted from quasars and other Active Galactic Nuclei (AGN) such as Seyferts, BL Lacs and Blazars
has been found to vary aperiodically on
time-scales from hours to decades and from X-ray to radio wavelengths. 
In the case of radio-loud and radio-quiet quasars optical flux changes over timescales as short as minutes and
with amplitudes ranging from a few hundredths to a few tenths of a magnitude 
(often
refereed to as `optical microvariability'),
have been widely reported (Miller, Carini $\&$ Goodrich \cite{mil89}, Stalin et al. \cite{stalin05}). The mechanism(s)
responsible for this short-term microvariability is still debated, with much of the visible spectral 
variability being explained by thermal processes associated with a black-hole accretion disk and/or non-thermal processes
associated with the relativistic beaming of a jet (Ramirez et al. \cite{ram10}, Pereyra et al. \cite{per06}).
In fact Sesar et al. (\cite{ses07})
have found that nearly all quasars vary at some level when observed over long enough times scales, with $\sim$ 90$\%$
of quasars found in the Sloan Digital Sky Survey (SDSS) varying by $>$ 0.03 mag over a six year time scale at
visible wavelengths. 
Although the actual cause of (long-term) variability in these objects is also still a matter of
intense debate, variability studies of these sources are important since (in principal) they can set limits on the size of the emitting
regions and can also constrain models of the emission mechanisms  thought to operate in these
active galaxies.

Numerous variability studies have been aimed at establishing correlations between
quasar visible variability and important physical parameters such as source luminosity, redshift, time-lag (i.e. the time period between
two sets of observations)
and radio loudness (Helfand et al. \cite{helf01}, Vanden Berk et al. \cite{van04}, de Vries et al. \cite{devries05}, Ramirez et al. \cite{ram09}).
The establishment of such correlations is important since there is increasing evidence that the source
of quasar variability
is linked to the accretion disk surrounding the central supermassive black
hole engine (Peterson et al. \cite{peter04}, Pereyra et al. \cite{per06}, Wold et al. \cite{wold07}). For example,
Bachev (\cite{bach09}) has studied the observed
time lags between the continuum emission from several optical bands for a large sample of quasars.
He demonstrates that the lags are broadly consistent with an emission model in
which the optical variations are largely due to the reprocessing of the
central X-ray radiation in a surrounding thin accretion disk. However,  both
Kelly et al.  (\cite{kelly09}) and MacLeod et al. (\cite{mac10}), using a `damped random walk model', have suggested
that these optical emission variations result from thermal
fluctuations in the accretion disk that are powered by an underlying stochastic process such as a turbulent magnetic field.
Clearly there is still much debate as to the interpretation of 
these optical variability studies (Kozlowski et al. \cite{koz10}), and other emission mechanisms such as
bursts of supernovae events close to the nucleus and extrinsic variations due to line-of-sight microlensing events
cannot  (as yet) be totally discounted (Ulrich et al. \cite{ulrich97}).Ê
It is mostly agreed that the level of quasar variability increases with decreasing rest wavelength, which is related to the
observation that their spectra become bluer in their bright phases. 
Several authors have claimed
a variability-redshift correlation in the visible, although the actual sign of the correlation in the
studies is debated (Giveon et al. \cite{giv99}, Trevese and Vagnetti  \cite{trev02}). For example,
Vanden Berk et al. (\cite{van04}) have claimed evidence for redshift evolution in quasar variability,
with greater variability occurring at higher redshifts. However, such a trend is only revealed
when an allowance is made for rest wavelength, absolute magnitude and time lag for each redshift interval.
Such corrections to the observed optical variability data are crucial, since spectral features may move in and
out of the photometric pass-bands with changes in the target redshift.

Quasars and AGN have also been well studied at ultraviolet (UV) wavelengths, such that typically their far UV spectra
appear as several strong emission lines (formed in gas located further away from
a central black hole and excited by continuum radiation) superposed on an underlying accretion-driven continuum.
In contrast, their near UV spectra contain only a UV continuum with a `bump' at 2800\AA\ superposed with the MgII
emission lines (Vanden Berk et al. \cite{vanden01}).
The UV variability of these objects has been widely reported,
particularly with reference to
reverberation mapping observations that in principal allow measurement
of the size of the broad emission-line region and can infer the mass of the alleged central black hole (Peterson
et al. \cite{peter93}, Kaspi et al. \cite{kaspi00, kaspi07}). UV photons are generally
believed to be radiated from the quasar supermassive black hole accretion disk when its gravitational
binding energy becomes released, while X-rays are produced in the disk corona.
A statistical survey of the UV variability of 93
quasar/AGN, as observed with
the I.U.E. satellite, was carried out by Paltani and Courvoisier (\cite{palt94}) who found no strong evidence
for correlations between variability and parameters such as redshift, luminosity or UV spectra-index. However,
changes and instabilities in the accretion disk must surely
be linked to their observed UV variability, but recognizing and understanding the
emission signatures from such features still remains problematic. 

To proceed further with studies of UV variability
we require access to far larger data sets than previously possible, and with the launch of the
NASA $\it Galaxy$ $\it Evolution$ $\it Explorer$ (GALEX) satellite (Martin et al. \cite{mar05}) we now have a
significantly large database
of far UV (FUV: 1350 - 1785\AA) and near UV (NUV: 1770 - 2830\AA) observations of quasars and AGNs that cover
a timespan of over 5 years.
Trammell et al. (\cite{tram07}) have 
presented a preliminary
analysis of the basic UV properties of  6371 SDSS Data Release 3 (DR3) selected quasars
observed during the first few years of the $\it GALEX$ mission. One major conclusion from this study
was that $>$ 80$\%$ of the DR3 listed quasars had a corresponding $\it GALEX$ NUV detection,
implying that many more quasars should be detected during the remaining years of operation
of the $\it GALEX$ satellite. Bianchi et al. (\cite{bianch09}) have shown that $\it GALEX$ is also
ideal for detecting quasars with blue (FUV-NUV) colors that are difficult to identify from visible observations alone.
Since the publication of these initial studies, newer and far
larger versions of both the
SDSS and $\it  GALEX$ catalogs are now in-hand.  In particular,  the 
catalog of quasars contained in the SDSS Data Release 5 (DR5) published
by Schneider et al.( \cite{schn07}) lists some 77,000+ objects (with
associated redshift and luminosity values), the majority of which
should have UV counterparts observable with $\it GALEX$. Furthermore, since
many quasars are known to be variable, this potentially enables us to study the UV
variability of a significantly large number of these sources.

The largest group of known variable sources listed in the $\it GALEX$ UV Variability-2 catalog are active
galaxies (Wheatley et al. \cite{wheat08}). This sample of sources was derived solely through selection due to their
observed UV variability, without prior knowledge that such sources were in fact AGN.   
These variable sources possessed redshifts as high as $\it z$ $\sim$ 2.5 and flux
changes as large as $\sim$ 2.3 AB mag. were found in both of the $\it GALEX$ near and far UV bands for
AGN that were repeatedly observed over a 3.5 year time period. Based on the apparent
quality of these data, we decided to
search the entire $\it GALEX$ archival database (which now spans 5+ years
of $\it GALEX$ observations) for spatial matches to quasars previously identified and listed in the SDSS DR5 catalog
and then, for those sources that had multiple visit measurements, determine which ones exhibited measurable
UV variability. This Paper discusses which $\it GALEX$ sources were identified with SDSS quasar counterparts,
which ones showed long-term  UV variability (as measured over timeframes $>$ 3 months) and which, if any, correlations
exist between this UV variability
and other astrophysically interesting source parameters.
 
\section{Data Analysis}
In order to produce a list of quasars that was suitable for an ultraviolet
variability study we first  searched the $\it GALEX$ GR5 data archival
release (at the Multi-Mission Archive at the Space Telescope Institute, MAST) for sources that 
spatially matched optically identified quasars listed in the SDSS DR5 spectroscopic catalog
of Schneider et al. (\cite{schn07}). A spatial match was deemed positive if the central position of the matching
$\it GALEX$ source was within $\pm$ 8 arc sec of the listed SDSS position, this number being slightly larger than the
positional error of the $\it GALEX$ observations for a point source detection.
This search resulted in 77, 249 quasars
that were common to both the SDSS and $\it GALEX$ ultraviolet source catalogs.

In order to produce a list of quasars whose UV variability could be accurately assessed as a function
of time, we then down-selected
the large list of UV-selected quasars using the following criteria. Firstly, we only selected quasars with either
(i) $\it GALEX$ observations in the All-sky Imaging Survey (AIS) that were brighter
than NUV$_{mag}$ = 19.0 and with an exposure time of at least 100s,  or (ii) those quasars with $\it GALEX$
Deep Imaging Survey (DIS) observations of 1500s duration or more (see Morrissey et al. \cite{mor07} for a detailed discussion
of the observational modes of the $\it GALEX$ satellite).
A further restriction was applied
such that only those quasars that had 5 or more observational
visits by $\it GALEX$ (and were detected in at least one of those visits) were selected. 
We note that an observational visit by $\it GALEX$ does not
necessarily result in a detection, since an intrinsically variable object may fall below the detection
limit throughout a particular visit.
We were guided in these choices by Figure 4
of Trammel et al. (\cite{tram07}) which shows the typical rms magnitude errors for an
AIS and DIS observation with $\it GALEX$. Those results demonstrate that the measurement error
for a typical AIS observation of a relatively UV bright (NUV$_{mag}$ = 19.0) object is approximately equal to
the measurement error for a 1550s DIS observation of a faint (NUV$_{mag}$ $>$ 21.0) source.
In Figure 1 we show a plot of
the number of these selected quasars as a function of the number of observations in
the $\it GALEX$ NUV and FUV channels. It can be seen that there is a significant number of sources that
have been visited between 5 and 20 times, with several sources being repeatedly observed
$>$ 100 times.
On application of the selection criteria of both exposure time and $\ge$5 visits,
this resulted in a list of
4360 quasars that were detected in at least one  observational visit by $\it GALEX$.

In Figure 2 we show a plot of these selected quasars as a function of the timespan
covering all of their 4360 NUV and 2709 FUV $\it GALEX$ observational visits (NB: we note
that the sample of FUV selected quasars is smaller than the number of NUV selected ones due to
operational constraints on the FUV detector that often precluded simultaneous FUV and NUV observations
of a source).
This Figure demonstrates
that our quasar sample is not biased towards
any particular timeframe of variability, with the majority of selected sources having been
repeatedly sampled over a 1 to 5 year timespan. We note that although the time between many observational
visits to a particular quasar may span several years, the target may be repeatedly observed on time-scales $<$ 1 day 
during a series of observational visits with $\it GALEX$.

All the data
shown in Figures 1 and 2 clearly indicate that our sample
of quasars is both large and contains significant numbers of
objects that have multiple ($\ge$ 5) observational visits spread over a significantly
long timespan. In order to ensure that quasars with potentially short-term variations of 
a few days had $\ge$ 5 observations, 
we have further restricted our target selection to those ones whose visits spanned a time-frame
$>$ 3 months. This selection criterion resulted in a final list of 3521 potentially UV variable quasars, of which 1942 had 10 or more
observational visits. To ensure that all of the UV-selected sources were not in any preferential direction on the sky,
in Figure 3 we show the galactic 
locations of our SDSS selected sample that have been observed 5 or more times. Since our selection
of targets was derived from the (northern hemisphere) SDSS survey catalog, clearly only
quasars with declinations $>$ -12$^{\circ}$
are available for present study. Although there are large areas of the sky where we have no UV-selected quasars,
Figure 3 shows no obvious sight-line preference apart from most sources being observed at high galactic
latitude away from the effects of absorption of ultraviolet flux by gas and dust in the galactic plane. 
We then searched this list of 3521 sources for the ones which possessed statistically
significant UV variability between each of their observational visits. Although the quasar nature of these sources at
visible wavelengths
was already known from the SDSS DR5 catalog, we decided that statistical confirmation of their associated UV
variability was also required for this study. Due to the limited sensitivity of the $\it GALEX$ observations this naturally restricted
our study to the more UV variable quasars in this sample.

\subsection{UV Variability Selection}
In order to determine which of the 3521 selected quasars were variable at least once during their series of observations, we require
knowledge of the size of the error associated with the estimation of each of the NUV and
FUV magnitudes measured for each visit to a particular source.
This error in this measurement is affected by the
length of the observation by $\it GALEX$, the position of the source on the UV detector and the associated
varying sky background signal which are both dependent on the (changing) orbital viewing characteristics of the
$\it GALEX$ satellite. Thus, in order to progress further in our study we clearly require a more consistent estimation
of the measurement error for the UV magnitudes derived from each observational visit by the $\it GALEX$ satellite.
Fortunately,
Gezari et al. (\cite{gez08}) have established a robust numerical method of assessing the statistical significance
of the variability of a series of $\it GALEX$ UV observations of the nuclei of normal quiescent galaxies,
which is directly applicable to our present QSO data set. The
method involves deriving the quantity, $\sigma$($<$m$>$), which is defined as the standard deviation ($\sigma$)
from the weighted mean ($<$m$>$) of a series of observed UV magnitudes (m).  For a series of observations
of the same object, the quantity of $\sigma$($<$m$>$) will vary such that
above some statistical level any observed change will be due to
intrinsic source variability. Gezari et al. (\cite{gez08}) estimated the Poisson errors for a series of
UV magnitude determinations using $\it GALEX$ (see equation (2) in that paper).  They
found that the measured standard deviation from a mean UV source magnitude
was dominated by systematic errors that were not accounted purely by the photometric error.
We note that the major contribution to the error in the estimation of a UV magnitude using $\it GALEX$ is thought to be due
to the (often varying) sky background level. Thus to be conservative,  Gezari et al. (\cite{gez08}) used a cut of 5-$\sigma$
above the mean magnitude level to determine intrinsic source variability. 

Values of $\sigma$($<$m$>$) for our quasar sample were also derived
from equation (2) of Gezari et al. (\cite{gez08}) for  each set of observational visits.
The values of NUV and FUV AB magnitudes recorded for each visit
to a selected quasar, together with their measurement errors, were retrieved from the
$\it GALEX$ GR5 data archival
release at the MAST. We used circular aperture magnitudes of a 6 arc sec radius, as listed in the MAST
catalog, which were obtained using the
$\it GALEX$ pipe-line generated SExtractor magnitude estimator (Bertin $\&$ Arnouts \cite{bert96}). 
In Figure 4 we show a plot of $\sigma$($<$m$>$) versus $<$m$>$ for the multiple NUV observations 
of 3521 of the selected quasars.  As one might expect, the value of  $\sigma$($<$m$>$), which is essentially a 
measure of the error in the derived mean AB magnitude of a quasar, increases with the faintness of the object under study.
All of these 3521 quasars were detected as being $\it optically$ variable in the SDSS DR5 catalog, and although they also all
may be UV variable (though this has never been actually proven in any large study), we have decided to take a
conservative approach in selecting only the more
 UV variable objects for further study in this paper. 
To this end we have decided to use a threshold cut using a standard deviation of 3-$\sigma$ above the level of $\sigma$($<$m$>$)
for a given mean magnitude ($<$m$>$) to reveal significant UV variability.  On application of this threshold cut we find
that the smallest detectable magnitude variation is 0.11 mag, which was determined for a 16.5 AB mag quasar of redshift $\it z$ = 0.4.

The 3-$\sigma$ threshold cut is shown superposed on both Figures 4 and 5 respectively for the NUV and FUV observed magnitudes.
Application of this statistical significance selection criterion resulted in the identification of 550 quasars as being NUV variable and
371 quasars as being FUV variable, with the vast majority of the FUV variable objects being common with those
identified as variable in the NUV. For simplicity, from now on we term this sub-set of selected variable objects as UVQs (i.e. ultraviolet
variable quasar stellar objects). The typical levels of observed UV variability for each of these UVQs,
can be best viewed when shown as NUV and FUV light-curves (i.e. the AB magnitude for each visit versus observation date).
In Figure 6 we show the FUV and NUV light-curves for a relatively bright (SDSS J163915.80+412833.6)
and a faint (SDSS J221629.33+002340) quasar, that were both repeatedly observed over a 4+ year time-frame with $\it GALEX$.
It should be noted that faint and variable objects can fall below the $\it GALEX$ threshold
of detection during a series of observations, and hence
the derived mean magnitude (which is computed from only actual detections within a series of visits) will
be of a higher value than had the object been detected during all visits. Thus, the computed change in magnitude,
which is a measure of the object's variability, will be underestimated. Finally we note that we did not discover any
UVQs that were highly variable in the FUV but not variable in the NUV.

\subsection{The Structure Function}
The mean variability of UVQs is often described by a Structure Function, which is the correlation between
two points in a set of measurements of an object's light curve (Simonetti et al. \cite{simon85}). Thus,
in its simplest form, it provides a statistical measure of  variability as a function of the time-lag
between many sets of observations.  However
we note,
as several authors have pointed out, there are limitations in its applicability to derive timescales
and amplitudes of quasar variability (Vio et al. \cite{vio92}). In addition, quasar variability
is most probably a complex function of several (possibly interdependent) physical parameters such as
luminosity, rest-frame wavelength, redshift, black hole mass and time-lag. 
Helfand et al. (\cite{helf01}) have pointed out that `the complex interdependence of the observed variability
(of quasars) on wavelength, redshift, bolometric luminosity, radio loudness and possibly other parameters makes it difficult
to draw definitive conclusions regarding the physics of the central engine or quasar evolution'. Thus, `caveat emptor'
should be the watchwords for any discussion of quasar variability dependencies, and thus we employ
a method by which
these dependencies of variability can be plausibly disentangled.

For our present study, in which most of the
identified UVQs each have $>$ 5 measurements, we start by using the standard formulation of
the Structure Function (SF) given by di Clemente et al. (\cite{diC96}):
\newline
\newline
	SF =  $\sqrt{\pi / 2 * \langle \mid \Delta m( \Delta \tau) \mid \rangle ^{2} - \langle \sigma_{n} ^{2} \rangle}$
\newline
\newline
where $\Delta$m($\Delta$$\tau$) is the difference in UV magnitude between any two observations
of a UVQ separated by $\Delta$$\tau$ in the quasar rest time frame, and $\sigma_{n}$$^{2}$ is the
square of the uncertainty in that difference (being equal to the sum of the two individual
observations' errors in quadrature). The final SF for all of our UVQ observations is just the rms of the summation
for all possible combinations of magnitude measurements for a given value of time-lag.

In Figure 7 we show log-log plots of the SF (i.e. the measure of UV variability) 
versus the rest frame time lag (in days) for both the NUV and FUV observations of each of the statistically significant variable UVQs
selected from Figures 4 and 5.
Following the work of Vanden Berk et al. (\cite{van04})
we have plotted the data averages in 10 time-lag bins,
chosen with equal intervals in the logarithmic rest-frame time lag. 
Note that the data as presented in Figure 7 make no account for any other plausible variability dependencies such as
luminosity, redshift or radio loudness. In this form, the variability function is often termed an `ensemble' Structure Function.
Error bars were determined by propagating the rms errors derived for the average
magnitude difference and measurement uncertainty in quadrature for each observation.

The data shown in Figure 7 reveal two important facets of the UV variability of quasars. Firstly,  the
amplitudes of variability in $\it both$ the NUV and FUV are far larger than those observed at visible wavelengths (which
are typically only $\sim$ 0.03 mag rms), and
secondly, the levels of FUV variability are greater than those observed in the NUV.

The NUV SF shown in Figure 7 increases linearly (with a slope of 0.439)
from a rest frame time-lag of 1 to $\sim$100 days, with a definite turn-over and subsequent
flattening for time-lags $\ga$ 300 days. The actual best-fit to all of the NUV data is a quadratic in log-log form.
A similar correlation of variability with time-lag has also been found in many other studies, and the flattening
of the curve at long time-lags has been reported by both Vanden Berk et al.( \cite{van04})
and Cristie et al. (\cite{crist97}). However we note that de Vries et al. (\cite{devries05}), in an optical survey of long-term quasar
variability up to timescales of $\sim$ 40 years,  has observed that this apparent
flattening of the SF only lasts for a few years before continuing its linear increase with time-lag. In
essence, they claim no significant turn-over of their SF,  such that visible quasar variability increases monotonically with increasingly
long time-lags. Although our sampling of UVQs is limited to time-scales of $<$ 5 years, inspection of the behavior of
the  NUV SF curve suggests that the rate of increase in UV variability as a function of time
decreases 
when observed over time-lags greater than $\sim$ 2 years.
The FUV Structure Function, also shown in Figure 7,  follows the same general trend as the NUV curve, except that we have observed
a greater
level of baseline variability of UVQs in the FUV. However, the change in FUV variability as a function of time-lag is less
pronounced than that in the NUV, as
indicated by the best-fit line of slope 0.292. The FUV data, like those recorded in the NUV,  also show a flattening in
UV variability for time-lags $\ga$ 300 days.

\section{Variability Correlation Plots}
Previous
studies of quasar variability in the rest-frame of the optical/UV regime have revealed several apparent physical correlations.
For example, trends of variability with wavelength and luminosity (first discovered by di Clemente et al. \cite {diC96} and
Hook et al. \cite{hook94}, respectively) have been
confirmed by Vanden Berk et al. (\cite{van04}) who have found that quasars are more variable at shorter rather
than longer wavelengths and the amplitude of variability decreases with increasing luminosity (i.e. intrinsically brighter
objects vary less). These results, and the SF method used to obtain them, have recently been 
verified for a sample of quasars observed over a 10 year period by MacLeod et al. (\cite{mac10}).
Furthermore, radio-loud quasars appear more variable than their radio-quiet counterparts (Helfand et al. \cite{helf01}),
and the apparent increase in variability for higher redshift quasars
is mainly due to the increase of variability with rest-frame frequency and to the increase of rest-frame frequency scanned at
high redshift for a given frequency (di Clemente et al. \cite{diC96}).
The Structure
Function derived in the previous section, which described UV variability as a function of rest-frame time-lag, took no account
of any of the other possible inter-dependencies such as luminosity or redshift. To reveal such multi inter-dependenices would require
4-dimensional plots, and thus we need an alternative method to explore other possible correlations with the UV SF.
Therefore, we closely follow the work of  Vanden Berk et al. (\cite{van04})  who have described a method that
 attempts to disentangle these possible physical
interdependencies.
Since each of the 10 time-lag bins of the UVQs 
plotted in Figure 7 contains
objects with a wide-range of other (differing) physical parameters, we start by subdividing the targets contained within each time-lag bin
into separate (and assumed to be degenerate) physical parameters. 
This then allows us to obtain a set of (quasi-independent) variability relations
with respect to the SF for a particular time-lag,  UVQ luminosity, redshift and radio loudness.  In the following subsections
we present these variability relationships, and then briefly discuss their possible astrophysical meaning in Section 4. 

\subsection{Time Lag - UV Variability Dependence}
Firstly we explore the dependence of UVQ NUV and FUV variability on the rest frame time-lag, in a form that
is (quasi) independent of the
parameters of luminosity and radio loudness. 
We divide all of the UVQs in our UV variability sample, whose redshift ($\it z$) values span
the range of 0 to 2.5, into three redshift bins of size 0 $<$ $\it z$ $<$ 0.47, 0.47 $<$ $\it z$ $<$ 1.33 and $\it z$ $>$ 1.33.
The physical relevance of these 3 redshift bins with respect to observations through the two $\it GALEX$ pass-bands
will be discussed in Section 4.
In Figures 8 and 9 we show log-log plots of the respective NUV and FUV Structure Functions versus rest frame time-lag,
with the data averages being plotted separately for each of these three redshift bins at any given rest frame time-lag value. 
Both of these plots, in
agreement with the simpler ensemble SFs shown in Figure 7, show that the amplitudes of both the NUV and FUV 
variability of UVQs increase as a function of rest frame time-lag, irrespective of the value of quasar redshift for time-lags $<$ 200 days.
For time-lags $>$ 300 days there is a pronounced rollover in the NUV SF for all redshift values, which is also observed
in the lower S/N FUV data. To further demonstrate this roll-over effect we have also included
 separate linear best-fits for each of the 3 redshift bin data sets in Figures 8 and 9.
These straight-line fits show that rate of increase of the NUV SF up to a time-lag of $\sim$ 200 days
 is greatest for sources of redshift 0.47 $<$ $\it z$ $<$ 1.33, an effect that is also mirrored in the FUV data set of
Figure 9. For time-lags $>$ 300 days it is clear that the same linear fits are inappropriate for all of the NUV and FUV data,
such that a far slower increase in the rate of the NUV and FUV SFs occurs.

Figures 8  and 9 both show that
for time-lags $<$ 20 days there is a definite trend for the higher-$\it z$ UVQs to be more NUV and FUV
variable than their low-$\it z$ counterparts. This trend has also been seen for visible quasars by de Vries et al. (\cite{devries05}).
However, for time-lags $>$ 100 days this trend mostly disappears such that
both the FUV and NUV SF variability amplitudes are similar for all values of UVQ redshift when observed over long time-lags.

\subsection{Luminosity - UV Variability Dependence}
Unlike redshift,  absolute luminosity is an intrinsic physical property of a quasar and thus an investigation of
UV variability (i.e. SF) as a function of this parameter should be less affected by other physical interdependencies.
We make use of the values of luminosity given in Schneider et al. (\cite{schn07}) for each of our
UV variable sources, and note that these
values were calculated using extinction corrected SDSS (absolute) $\it i$ magnitudes and with
the assumption that the quasar spectral energy distribution
in the ultraviolet-optical can be represented by a power law. 

We start by separating the possible dependency effects of 
rest frame time-lag by dividing the data into four time-lag bins of 1 - 10 days, 11 - 100 days,  101 - 500 days and $>$ 500 days. 
In Figure 10 we show log-log plots of both the NUV and FUV variability amplitudes (i.e. the SF's)
as a function of (averaged) absolute magnitude in bins of 2 mag width
for each of the aforementioned time-lag bins. The best-fits to these data show that there is a slight decrease in both
the NUV and FUV SFs as a function of increasing luminosity for time-lags $>$ 100 days. Similar (visible) luminosity plots have
also been constructed by Vanden Berk et al. (\cite{van04}) for a sample of 25,000
quasars in the SDSS survey, in which they similarly found that intrinsically brighter objects vary less than
intrinsically fainter ones. However, for shorter time-lags we find the opposite behavior in which
the NUV and FUV SFs increase as a function of increasing luminosity (we note the apparent exception of the NUV SF
for time-lags $<$ 10 days, but this may be due to small number statistics with associated larger errors).
The NUV and FUV data shown in Figure 10
also show that variability amplitudes are greater at longer time-lags than at shorter time-lags for all
values of UVQ luminosity. This behavior is also revealed in the SFs of Figures 7 - 9, which were derived 
independently of luminosity.
This difference in variability for both short and long time-lag values is discussed further in Section 4.1 with
respect to the work of Kelly et al. (\cite{kelly09}).

We note that there are no low-luminosity UVQs detected at high redshift in our sample, such that the low
luminosity bins in both plots of Figure 10 are populated mostly by low-$\it z$ objects.
Therefore, since the data in Figure 10 may still contain some hidden dependency on quasar redshift, in Figure 11 we re-plot  the
NUV and FUV variability data but now
subdivided into bins of  redshift. For convenience of plotting we have selected representative
redshift bins of 0.5 $<$ $\it z$ $<$ 0.895 and $\it z$ $>$ 2.028 for the NUV data, and bins of 0.5 $<$ $\it z$ $<$ 0.895
and 1.4 $<$ $\it z$ $<$ 2.03 for the FUV data, all plotted for the same two representative time-lag
bins as used in Figure 10. For the NUV data in Figure 11 the 4 curves of SF versus luminosity do not
precisely follow the same trends as shown in Figure 10, although there is an overall general trend
of decreasing SF value with increasing luminosity. This general trend of decreasing variability
amplitude is also observed in the FUV data, except for the case of
high $\it z$ and small time-lags where an increase of SF variability with luminosity is observed. Since the general trends
seen in both Figures 10 and 11 are similar, 
we can deduce that redshift does not seem to have a profound effect on
the UV variability- luminosity nature of these objects, and also that the shape of the UV SFs is only weakly dependent on
quasar luminosity.

\subsection{Radio Loudness - Variability Dependence}
In Figures 12 and 13 we
plot the SFs for the NUV and FUV variability of our quasar sample against their rest frame time-lag for both
radio loud and radio faint UVQs.  
Based on their peak 20cm radio flux density as measured in the VLA FIRST survey 
and using the absolute SDSS $\it i$ magnitudes and distances listed in Schneider et al. (\cite{schn07}) to derive a measure
of their optical luminosity, we compute values of the ratio of radio to optical luminosity in the rest frame (R$^{*}$). Using
equations (1) and (2) of Sramek $\&$ Weedman (\cite{sram80}) to derive R$^{*}$, we use the widely accepted definition of R$^{*}$ $<$ 1
for radio quiet sources and R$^{*}$ $>$ 1 for radio loud sources. Since many of our variable quasars do not have
listed values for their radio fluxes, our NUV sample is restricted to 484 radio loud and 66 radio quiet sources and
our FUV sample contains  327 radio loud and 44 radio quiet sources. Hence, the reader is reminded that we
are dealing with small number statistics for the radio quiet source data in both Figures 12 and 13.
 
Firstly, inspection of both Figures clearly shows that the overall level of SF variability is greater in the FUV than the NUV for
both radio loud and radio quiet UVQs. This is merely a re-statement of the behavior of the entire UVQ sample, as shown
in Figure 7. However Figure 12 also indicates that radio loud sources are $\sim$ a factor 2 more NUV variable as a
 function of time-lag than radio quiet sources,
and the plot also shows a larger scatter in
the shape of the NUV SF for radio loud than for radio quiet UVQs. However, there is only a marginal difference between the best-fit slopes
to these NUV SF data for both types of radio source.
Therefore, our data would appear to support the notion that NUV variability is not strongly linked to
the radio properties of UVQs, and as such,  the physical mechanism 
responsible for the NUV variability of radio detectable UVQs operates irrespective of their radio flux.

The FUV data shown in Figure 13 have far fewer measurements and larger errors than the NUV data, but again the radio loud
UVQs appear to be more variable as a function of time-lag (especially for periods $>$ 100 days) than the radio quiet sources.
The best-fit slope of 0.16 to the FUV radio loud data is significantly different from that of the best-fit slope of 0.059 for
the radio quiet data. This FUV variability difference is most pronounced for time-lags $>$ 100 days. However, we note
that this conclusion is based on only a small number of UVQs. We note that several authors 
have claimed that radio-loud quasars are (marginally)
more variable than their radio-quiet
counterparts (Eggers et al. \cite{egg00}, Helfand et al. \cite{helf01}), whereas 
Bauer et al. (\cite{bauer09}) have found no evidence for any change in the optical variability amplitude of quasars
with increasing radio loudness (luminosity). Several studies have found that optical microvariability, operating
over time-scales of minutes to hours, occurs in quasars regardless of their radio properties (Ramirez et al. \cite{ram09}).
Our present NUV (and FUV) observations support the notion in
 which radio loud UVQs are only marginally more UV variable than radio quiet sources, irrespective of the time-frame of observation.

\section{Summary Discussion}
Before briefly discussing any physical interpretation of the correlations shown in Figures 7 - 13
of Section 3, 
we first provide a short summary of the underlying UV properties of the UVQ sample.
We also remind the reader that due to our conservative approach in the selection of UV
variable quasars, these findings may only apply to the more UV variable of this class of object.
In Figure 14 we show a typical quasar
rest-frame spectrum (adapted from Vanden Berk et al. {\cite{vanden01}), which
shows the prominent Ly$\alpha$, CIV, CIII and MgII emission lines superposed on
a significant underlying UV continuum level, which for wavelengths blueward of Ly$\alpha$ is heavily
extinguished by the Ly$\alpha$ forest of absorption lines. The entire UV energy distribution is often referred to as
the `big blue bump', and is
thought to be dominated by the 10$^{4-5}$K thermal emission from a central black hole accretion disk.
Based on the sensitivity and observational strategy of the $\it GALEX$ satellite,
the great majority of quasars that can be detected possess redshifts
in the range 0 $\le$ $\it z$ $\le$ 2.5 (Bianchi et al. \cite{bianch07}).
For quasars with redshifts in
the range from $\it z$ = 0.11 to 0.47 the strong Ly$\alpha$ emission line is present within the $\it GALEX$
FUV passband, and for redshifts of $\it z$ = 0.46 to 1.33 the Ly$\alpha$ line is present within the NUV passband. 
In addition, the Ly$\alpha$ forest of absorption lines (which considerably reduces the UV flux in a quasar spectrum)
affects the entire $\it GALEX$ passbands for redshifts of $\it z$ $>$ 1.33. Therefore, any observed UV variability within either of
the $\it GALEX$ FUV or NUV passbands will 
be highly dependent on which of the emission/absorption/continuum features present in the
quasar spectral energy distribution have been redshifted into, or out of, those two passbands and also which
of these features are deemed to be most responsible for the physical origin of the observed UV variability.
Therefore, our selection of 3 redshift bins (0.11 $<$ $\it z$ $<$ 0.47, 0.47 $< $ $\it z$ $<$ 1.33 and $\it z$ $>$ 1.33) for
the plotting of data in Figures 8 to 9
represent the appearance and disappearance of the strong Ly-$\alpha$ line in the FUV and NUV bands.
Additionally, because of time dilation effects, high redshift objects are
potentially observed for less time than low redshift ones. This will possibly affect any study of the time-lag
on UVQ variability, which is why we have binned our time-lag data in the rest frame.

Presently, explanations of the primary cause of quasar variability can be broadly split into 3 classes: (i) gravitational microlensing,
(ii) discrete-event (Poissonian) processes and (iii) accretion disk instability models. Vanden Berk et al. (\cite{van04}) have
discussed these various processes (see the many references therein) and their 
visible data favor an accretion disk instability (or jet) cause of the observed optical variability. 
Since many of our
present results (shown in Figures 7 - 13) generally agree with the visible observations of quasar variability
by Vanden Berk et al. (\cite{van04}), we shall only address whether a model
involving accretion disk instabilities may explain our UV variability data.
However, the reader should note that no 
theoretical model, irrespective of its underlying physical mechanism, currently exists that
can be directly compared with their (or our own UV) observations. 

Long-term (7.5 yr) changes in the UV continuum
level of order 25$\%$ - 125$\%$ have been reported for low luminosity quasars (Kaspi et al. \cite{kaspi00}), which is about 
twice the continuum variability observed over a similar period for high luminosity (high-$\it z$) objects (Kaspi et al. \cite{kaspi07}).
Interestingly for this latter sample of (11) quasars, none showed Ly-$\alpha$ emission variations to a level of $<$ 7$\%$ 
over the 6-year period of observations. A similar case of non-variability of the Ly-$\alpha$ line has also been reported
for observations of  3C 273 recorded over a 15yr timeframe by Ulrich et al (\cite{ulrich93}). 
Several authors have claimed that
quasar continuum variations are dominated by processes
that are intrinsic to the accretion disk (Peterson et al. \cite{peter04}), and Trevese, Kron $\&$ Bunone (\cite{trev01})
have suggested that the observed spectral changes are consistent with temperature variations
of a blackbody of $\sim$ 2.5 x 10$^{4}$K that emits in the UV. More recently Pereyra et al. (\cite{per06})
have also found
that most of the optical-UV variability observed in quasars is due to changes in the UV continuum level which
can be directly attributed to
processes involving changes in the disk mass accretion rates. Repeatedly visible spectroscopic
observations of 2,500 quasars over $>$ 50 day time-frame by Wilhite et al. (\cite{wil05}) have shown that
the strongest quasar emission lines vary by only 30$\%$ as much as the underlying continuum level.
All these results are therefore consistent with the increasingly
widely-held belief that
the majority of a quasar's UV-optical spectrum arises directly from the accretion disk and
that changes
in the level of the underlying UV continua of quasars are probably the main driver of the observed variability
in these objects.

\subsection{Ensemble Structure Function Dependence}
The ensemble Structure Functions
shown in Figure 7 clearly reveal that both the FUV and NUV variability of UVQs are well correlated with rest frame time-lags
of
$\la$ 300 days, such that the UV variability amplitude increases almost linearly with increasing time-lag. However,
for longer time-lags there is a rollover in both SF curves, which is
also observed at visible wavelengths. It is therefore initially tempting to attribute this
rollover to a change in the underlying physical process that cause the observed UV variability.
However, a rollover in the SF at large time-lags
has recently been fully explained by the `damped random walk' model of quasar variability
presented by Kelly et al. (\cite{kelly09}) and
MacLeod et al. (\cite{mac10}).  These authors model quasar light curves as a continuous time stochastic process, which results
in the estimation of a characteristic timescale and amplitude for visible flux variations. In the visible, these
flux variations are thought to result from thermal fluctuations that are driven by an underlying stochastic
process such as a turbulent magnetic field associated with the black hole accretion disk.
The shape of their
power spectrum (i.e. SF) is predicted to be 1 / f$^{2}$, which then flattens to white noise
at timescales long compared to the relaxation time.  This relaxation time is interpreted (for quasar flux variations) as
the time required for the time series of fluctuations to become roughly uncorrelated. From our present NUV and FUV data
this relaxation time is $\sim$ 400 days.

The reader should be reminded, however, that our $\it GALEX$ observations
are biased towards the detection of brighter and lower-redshift objects which could potentially be making a larger
contribution to the ensemble SF at long time lags compared with the higher redshift (and fainter) UVQs. Hence, although
the ensemble SFs shown in Figure 7 are important indicators of quasar variability,
only by discussing the potential dependencies shown Figures 8 - 13 can we better interpret these data.

\subsection{Time Lag Dependence}
The NUV and FUV SFs shown in Figures 8 and 9 as a function of time-lag follow the same general trend
as their ensemble SF counterparts shown in Figure 7, in that the amplitude of 
UV variability increases (linearly) as a function of rest frame time-lag
up to $\sim$ 200 days and thereafter there is a pronounced rollover in the NUV and FUV SFs for all
redshift values. This overall trend is similar, irrespective of the value of UVQ redshift.
For short time-lags $<$ 20 days there is a definite trend for higher-$\it z$ UVQs to be more NUV and
FUV variable than
their low-$\it z$ counterparts, but this trend all but disappears for time-lags in excess of 100 days.
If we assume that the observed variability is mainly due to changes in UV line emission,
then we might expect the NUV band to exhibit greater variability than that recorded in the FUV band
for objects with $\it z$ = 0.46 to 1.33 (i.e. the range in which the strong Ly-$\alpha$ line
enters the NUV band pass). However, our observations do not strongly support this notion since the
SF variability levels of the FUV data for these objects are similar to those of their NUV variability.
Thus, it seems likely that
the observed FUV variability for the UVQs in this redshift range may be better explained through (stochastic)
changes in the underlying UV continuum
level. 

\subsection{Luminosity Dependence}
Although it is
generally agreed that the intrinsic luminosity of a quasar may be linked to matter accreting from a disk
onto a black hole, the physical mechanism(s) responsible for the observed changes
in variability are still widely debated. As pointed out in
the Introduction, possible variability mechanisms from extrinsic gravitational sight-line microlensing
to intrinsic accretion disk instabilities have been forwarded to explain the observations.
Using $\it I.U.E.$ satellite observations, Paltani $\&$ Courvoisier (\cite{palt97}) found that 
quasar variability and luminosity (absolute magnitude) were related by
a power law slope of -0.08, such that relative variability decreases as the
quasar luminosity increases. This anticorrelation between quasar variability and luminosity has also
been confirmed at optical wavelengths using large 
homogeneous data sets (Bauer et al. \cite{bauer09}, Vanden Berk et al. \cite{van04}). However, we note
that Ai et al. (\cite{ai10}) claim that the inverse dependence of variability on luminosity is not significant
for a sample of SDSS selected AGN (mostly quasars). They claim that their
analysis was capable of recovering AGN
with a small variability in the low luminosity
regime which were largely missed in most of the
previous studies. Furthermore, previous
relationships between quasar luminosity and other parameters
such as black hole mass and spectral line-widths have mostly been derived from single-epoch estimates and using (mostly untested)
extrapolations from low  to high redshift objects (Vestergaard $\&$ Peterson \cite{vest06}). 

Our present findings shown in Figure 10 indicate that there is a general decrease in both NUV and FUV
variability as a function of increasing absolute magnitude for time-lags $>$ 100d. The opposite
 effect is observed for time-lags $<$ 100d (apart from the weak exception in the NUV for time-lags
$<$ 10d). Figure 11 shows that this general decrease in UV variability as a function of luminosity
is not strongly coupled to the UVQ redshift, in conflict with other (visible) studies that have claimed a redshift-variability
relationship (Trevese $\&$ Vagnetti \cite{trev02}). Both of the NUV and FUV SFs shown in Figure 10
reveal a similar behavior to those shown in Figures 7 -9 (which were derived 
independently of luminosity), 
in that variability amplitudes are greater at longer time-lags than
for  shorter time-lags for all
values of QSO luminosity. Therefore, it would seem from our present data that the intrinsic luminosity of a UVQ  may only
be weakly connected to the
origin of the physical mechanism causing the UV variability, and that this mechanism operates over most time-lags
and is not strongly dependent of the UVQ redshift. As discussed in Section 4.1, although the actual underlying physical
causes have yet to be proven, the application of  a `damped random walk'  model of stochastic thermal fluctuations
from an accretion disk can statistically explain the general shape of the ensemble variability
of quasars, and in particular the shape of their SFs as a function of time-lag (Kelly et al. \cite{kelly09},
MacLeod et al. \cite{mac10}). Therefore, in order to explain the majority of our UV variability data we do not have to invoke various
physical mechanisms that operate over different time-lags that depend on
various physical parameters, but instead the gross behavior of the UV variability of quasars can
essentially be described by a 1 / f$^{2}$ dependence that flattens to a near constant at low frequencies
(i.e. long time-lags). This seems to be true apart from the effect noted in Section 4.2 (and shown
in Figures 8 and 9) in which higher-$\it z$
UVQs appear to be more variable than their low-$\it z$ counterparts for time-lags $<$ 20 days. Whatever
physical (or observational) effect is responsible for this behavior essentially disappears from our data for time-lags $>$
100 days.

\subsection{Radio Loudness - UV variability}
Although the data presented in both Figures 12 and 13 have been constructed with relatively few UVQs, 
we note the near similarity between the shape of the NUV SFs for both radio loud and radio quiet objects.
In addition, we note that both Figures also show that radio loud UVQs are marginally more variable than radio-quiet
sources (a factor 2 in the NUV and a factor 1 - 3 in the FUV, depending on time-lag), in agreement with the work
of both Eggers et al. (\cite{egg00}) and Helfand et al.( \cite{helf01}). Similarly,
several other visible variability studies (Vanden Berk et al. \cite{van04},
Ramirez et al. \cite{ram09}) have revealed a correlation between radio loudness and UV/optical variability amplitude that
was suggestive but not conclusive. The source of quasar radio emission
is thought to originate either in an accretion disk around the central black hole or in shock waves in a relativistic
jet ejected from a region around the central black hole. Since the processes associated
with a powerful nuclear radio source probably influence UV variability over time-lags
$\la$ 100 days, we might expect the greatest difference in variability between radio loud and radio quiet objects to
appear in this time-regime. Both of our FUV and NUV data do not support such a scenario, and to the contrary the (sparsely sampled)
FUV SF suggests that the greatest variability difference occurs for time lags $>$ 100 days for radio sources.
Thus, although our data do reveal marginal differences between the behavior of radio loud and radio quiet sources,
we believe that the radio properties of UVQs appear only play a minor role in the gross levels of observed UV variability.

\section{Conclusion}
Using archival observations
recorded at NUV and FUV wavelengths over a 5+ year timeframe with the NASA
Galaxy Evolution Explorer ($\it GALEX$) satellite, we have performed an ultraviolet
variability analysis of quasars that have optical counterparts in the
Sloan Digital Sky Survey (SDSS) DR5 spectroscopic catalog of Schneider et al. (2007).  The application of a set of
(conservative) selection criteria has resulted in the identification of 550 quasars in the NUV and 371 in the FUV as being 
statistically significantly variable using data recorded in 5 or more $\it GALEX$ observations
spread over time periods $>$ 3 months. Typically we have observed a wide variety of UV variability levels ranging from
0.11 to 3.0 AB magnitudes arising in quasars with redshifts of 0 $<$ $\it z$ $<$ 2.5.

In order to explore possible correlations within these variability data we have used a Structure Function analysis
that closely follows the work of Vanden Berk et al. \cite{van04}, who performed a similar analysis on
visible observations of SDSS quasars. By subdividing the variable quasars into time-lag and/or redshift bins we have been
able to obtain a set of (quasi-independent) variability relations with respect to the Structure Function (SF) for values of
rest frame time-lag, luminosity and radio loudness. Since our selection criteria for UV variability was quite
conservative, our analysis of the SFs may only be applicable to the more UV variable QSOs.

The SF - time-lag plots clearly reveal that the
amplitudes of variability in both the NUV and FUV are far larger than those observed in quasars at visible wavelengths.
Additionally, the levels of FUV variability are greater than those observed in the NUV at any given value of time-lag.
We also find that the amplitudes of both the NUV and FUV variability of UVQs increase as a function of rest frame time-lag, irrespective of the
value of quasar redshift, for time-lags $<$ 200 days. For time-lags $>$ 300 days there is a pronounced rollover in the NUV SF for all
redshift values, which is also observed (with a lower significance) in the FUV variability data. These data also show
that for time-lags $<$ 20 days there is an apparent trend for higher-$\it z$ UVQs to be more NUV and FUV variable
than their low-$\it z$ counterparts. However, none of of other plots of SF dependence show any strong dependence
on redshift, and this effect may be due to some other physical dependency. Overall, these results are
best interpreted under the assumption that the UV variability is linked to variations in the level of continuum, 
and not line, emission.

Our data also indicate that for time-lags $>$ 100 days the more luminous UVQs tend to be less NUV
and FUV variable, whereas the opposite effect is mostly found for shorter time-lags (with the weak exception
in the NUV for lags of $<$ 10 days).
This difference in the level of SF
variability between long and short time-lags is also seen in the ensemble SF - time lag plots,
which were derived 
independently of luminosity. 
We also find that UVQ redshift
does not seem to have a profound effect on any of the variability-luminosity relations (alhtough
we note the previously mentioned trend for higher-$\it z$ UVQs to be more NUV and FUV variable
than their low-$\it z$ counterparts as a function of short time-lags).
The apparent weak effect of redshift on the UV variability-luminosity
 is perhaps not
surprising since luminosity is an intrinsic physical property whereas redshift is not. These data can possibly be 
interpreted through the application of  a `damped random walk'  model of stochastic thermal fluctuations
from an accretion disk which can statistically explain the general shape of our VUV SFs ((Kelly et al. \cite{kelly09},
MacLeod et al. \cite{mac10}).

Although our data set is small, it seems that radio loud sources are marginally more NUV variable as a function of time-lag than radio quiet UVQs.
Therefore it would seem that whatever the main mechanism is responsible for the observed NUV variability in quasars, it seems
to operate irrespective of their radio flux. Such behavior can also be explained by the damped random
walk model. Our data also hints that for time-lags $>$ 100 days radio loud sources may be
more FUV variable than radio quiet sources, although this conclusion is based on small number statistics.

In summation, our analysis favors a quasar model in which UV variability is mainly due to (stochastic) changes in the underlying
continuum level, rather than models that favor gravitational microlensing or discrete-event processes. Finally, we note
that as yet, no detailed quasar model currently exists that
can be directly compared with our observations of NUV and FUV variability.
   
\begin{acknowledgements}
We particularly acknowledge the dedicated work of the $\it GALEX$ mission operations support staff at JPL/Caltech in 
Pasadena. Financial support for this research was provided by the NASA $\it GALEX$ Guest Investigator program
and NASA grant NAS5-98034 to UC Berkeley. We are indebted to Dr. Suvi Gezari who made many
 suggestions that have greatly improved this paper.

\end{acknowledgements}

\begin{figure}
\center
{\includegraphics[height=9cm]{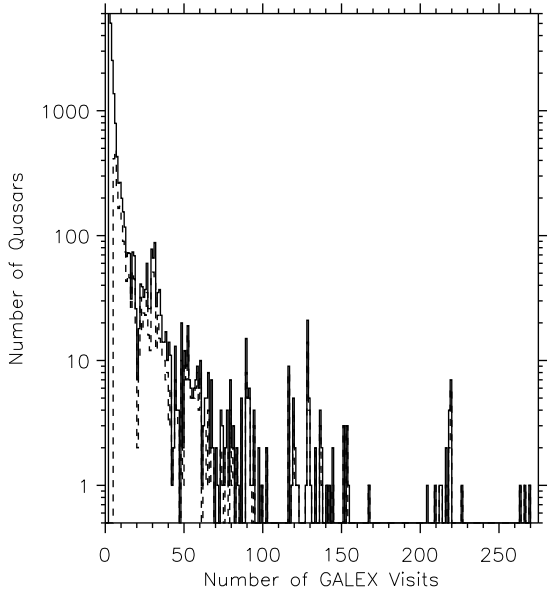}}
\caption{Plot of the number of $\it GALEX$ repeated visits for quasars with DIS integration times of $>$ 1000s and bright quasars with AIS observation of $>$ 100s (see text for selection criteria). Full line is for quasars with $\it GALEX$ NUV observations and dashed line is for quasars with FUV observations.}
\label{Figure 1}
\end{figure}

\begin{figure}
\center
{\includegraphics[height=9cm]{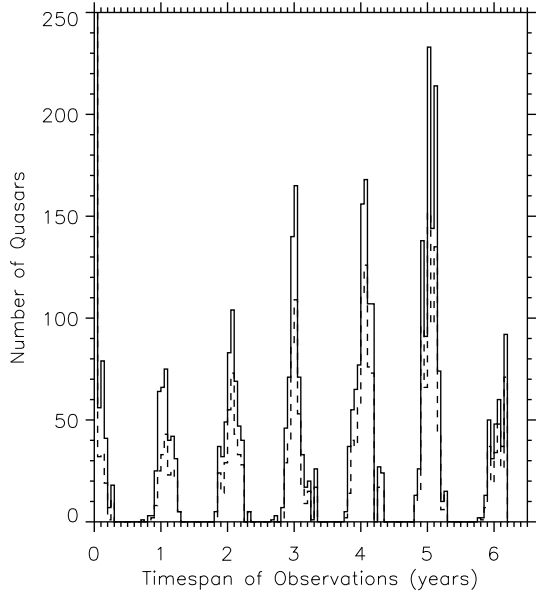}}
\caption{Plot of the number UV selected quasars and the time period spanning all of their $\it GALEX$ observations.
Full line is for quasars observed in the $\it GALEX$ NUV pass-band, dashed line is for quasars observed in the FUV pass-band.}
\label{Figure 2}
\end{figure}

\begin{figure}
\center
{\includegraphics[height=10cm]{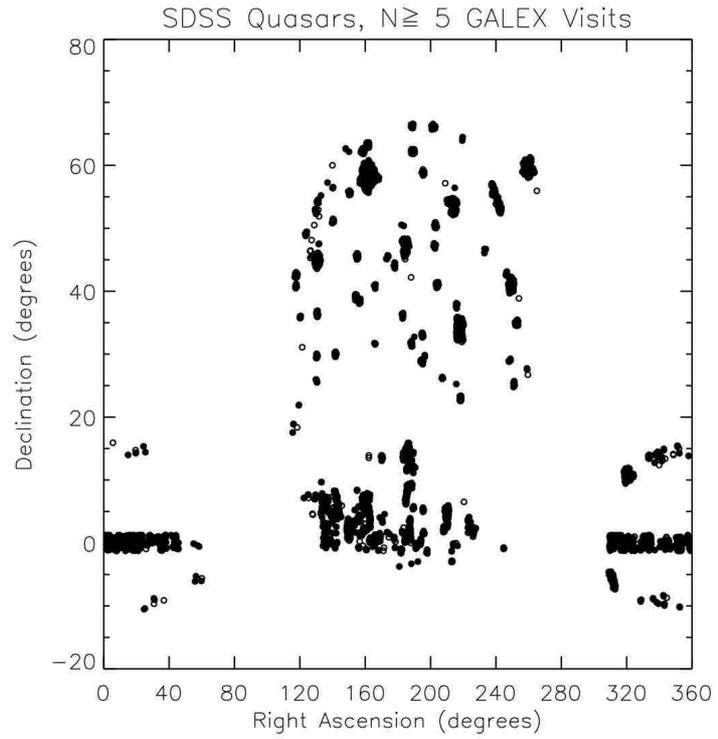}}
\caption{Galactic distribution of the UV selected quasars with $\ge$ 5 observational visits by $\it GALEX$. Filled circles are quasars with both NUV and FUV observations, while open circles are quasars with NUV only observations by $\it GALEX$.}
\label{Figure 3}
\end{figure}

\begin{figure}
\center
{\includegraphics[height=10cm]{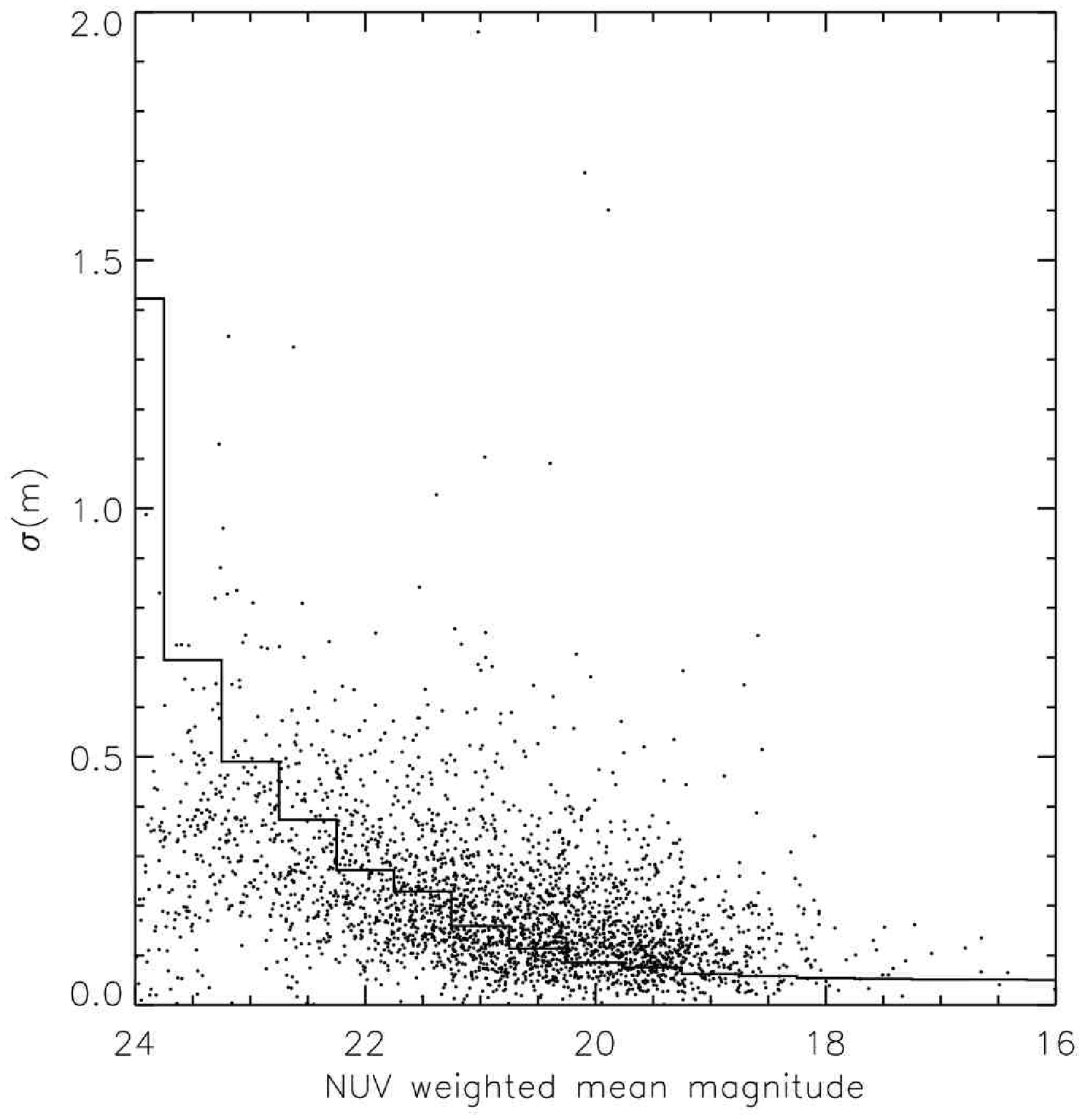}}
\caption{Plot of the standard deviation ($\sigma$) from the weighted mean ($<$ m $>$) of a series of observed NUV magnitudes (m) for
the sample of quasars with $\ge$5 observational visits by $\it GALEX$. Superposed on this plot (full line) is the theoretical 3-$\sigma$
variance from the measured mean NUV magnitude. Only points above this variance were selected as being statistically variable in the NUV.}
\label{Figure 4}
\end{figure}

\begin{figure}
\center
{\includegraphics[height=8cm]{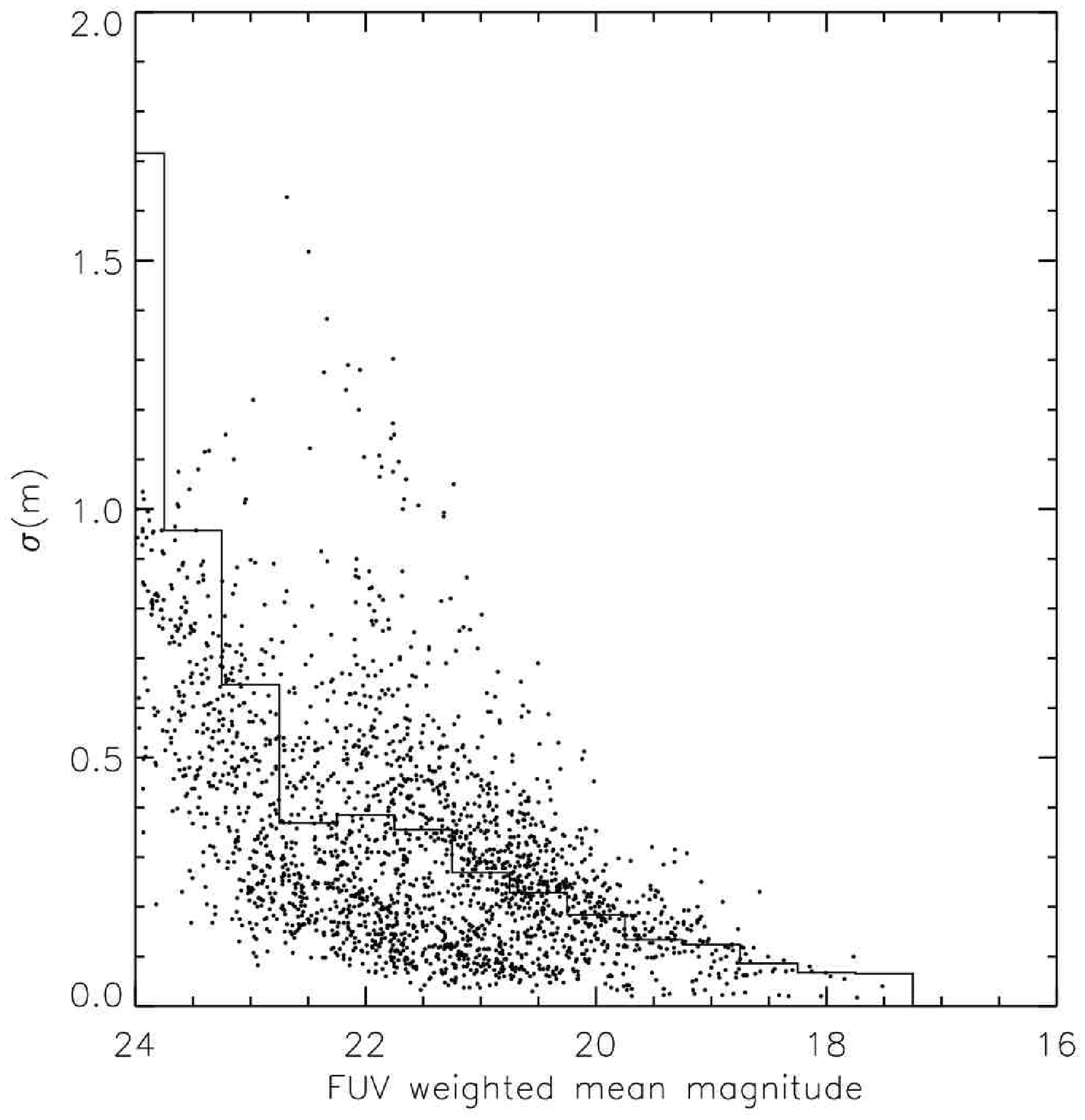}}
\caption{Plot of the standard deviation ($\sigma$) from the weighted mean ($<$ m $>$) of a series of observed FUV magnitudes (m) for
the sample of quasars with $\ge$5 observational visits by $\it GALEX$. Superposed on this plot (full line) is the theoretical 3-$\sigma$
variance from the measured mean FUV magnitude. Only points above this variance were selected as being statistically variable in the FUV.}
\label{Figure 5}
\end{figure}

\begin{figure}
\center
{\includegraphics[height=8cm]{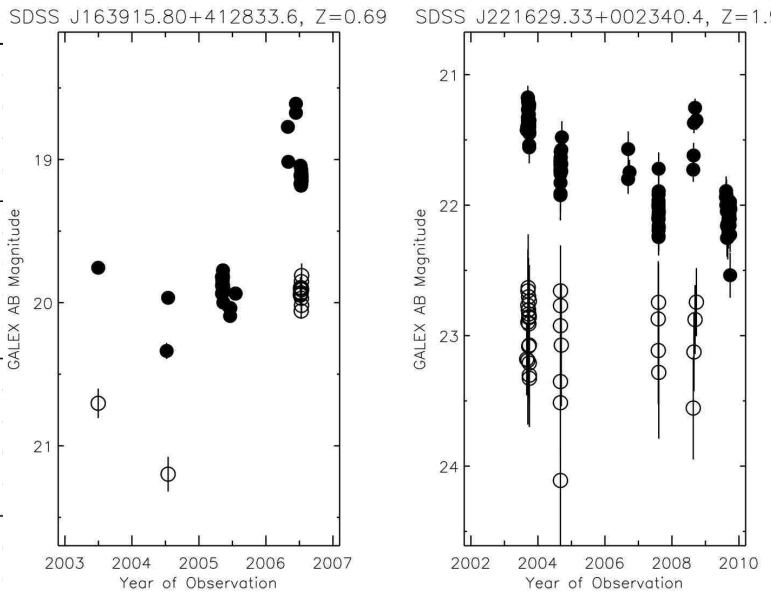}}
\caption{NUV (filled circles) and FUV (open circles) light curves for the UV-variable quasars (UVQs) SDSS J163915.80+412833.6 (left) and SDSS J221629.33+002340 (right) repeatedly observed over several years with the $\it GALEX$ satellite. Error bars represent the uncertainty in the measured AB magnitude (see text).}
\label{Figure 6}
\end{figure}

\begin{figure}
\center
{\includegraphics[height=10cm]{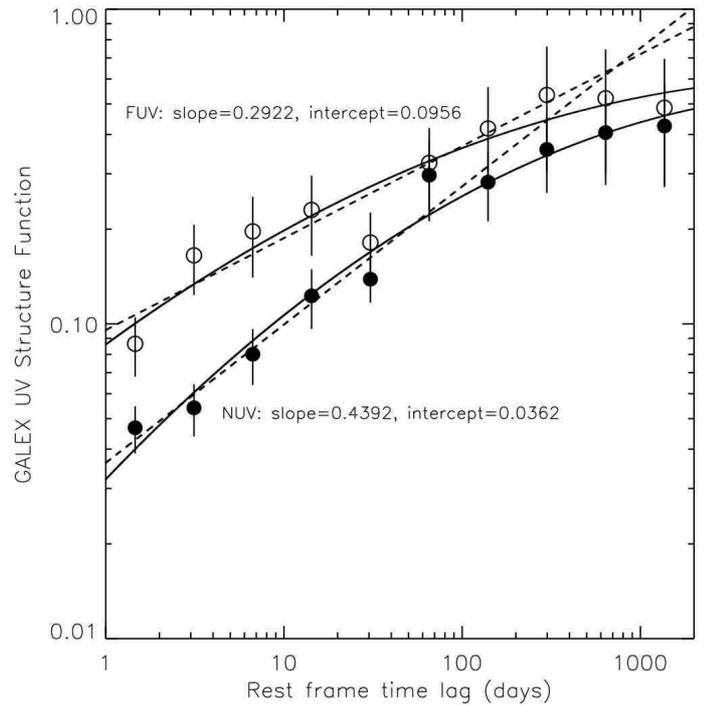}}
\caption{Log-Log plots of the NUV (filled circles) and FUV (open circles) structure functions with respect to the rest frame time lags for the UV-variable quasars. The data averages are shown in 10 bins of equal logarithmically spaced time lags. Best fit straight lines (dashed) and quadratic fits (full line) are shown for both Structure Functions. Both sets of data reveal a flattening of their structure functions for time lags $>$ 300 days.}
\label{Figure 7}
\end{figure}

\begin{figure}
\center
{\includegraphics[height=8cm]{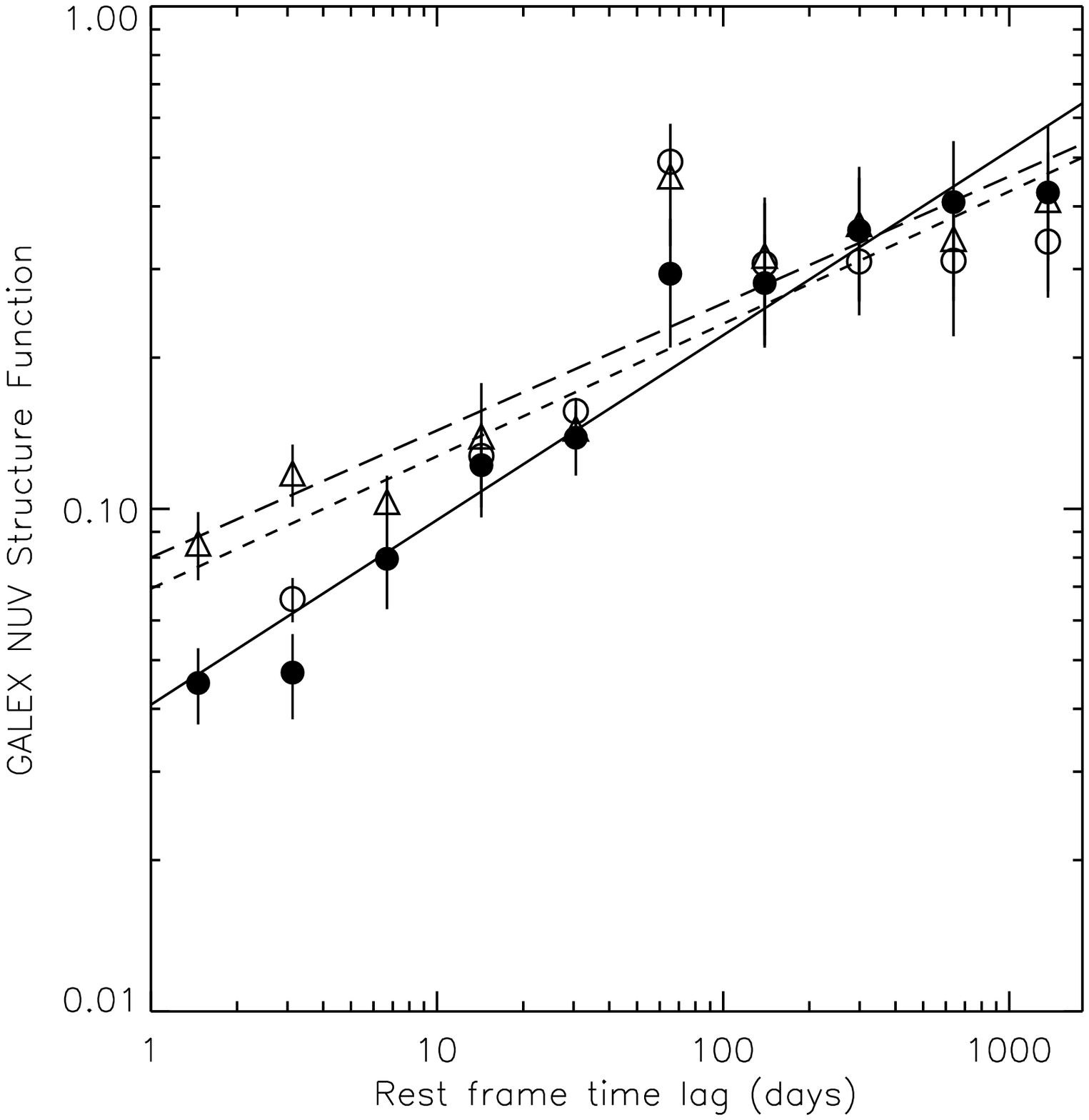}}
\caption{Log-Log plots of the NUV structure function (separated into 3 redshift bins) with respect to the rest frame time lags for the UV variable quasars. The data are shown for the same 10 bins of time lags as Figure 7, but with the data averages plotted for redshift bins of size 0  $<$ $\it z$ $<$ 0.47 (open circles),  0.47 $<$ $\it z$ $<$ 1.33 (filled circles) and $\it z$ $>$ 1.33 (open triangles). The vertical lines represent error bars for the binned data points. Linear best-fits are also shown for each of the redshift bin data sets; long dashed line (open circle data), full line (filled circle data) and short dashed line (open triangle data). None of these linear fits match the NUV data for time-lags $>$ 300 days.}
\label{Figure 8}
\end{figure}

\begin{figure}
\center
{\includegraphics[height=8cm]{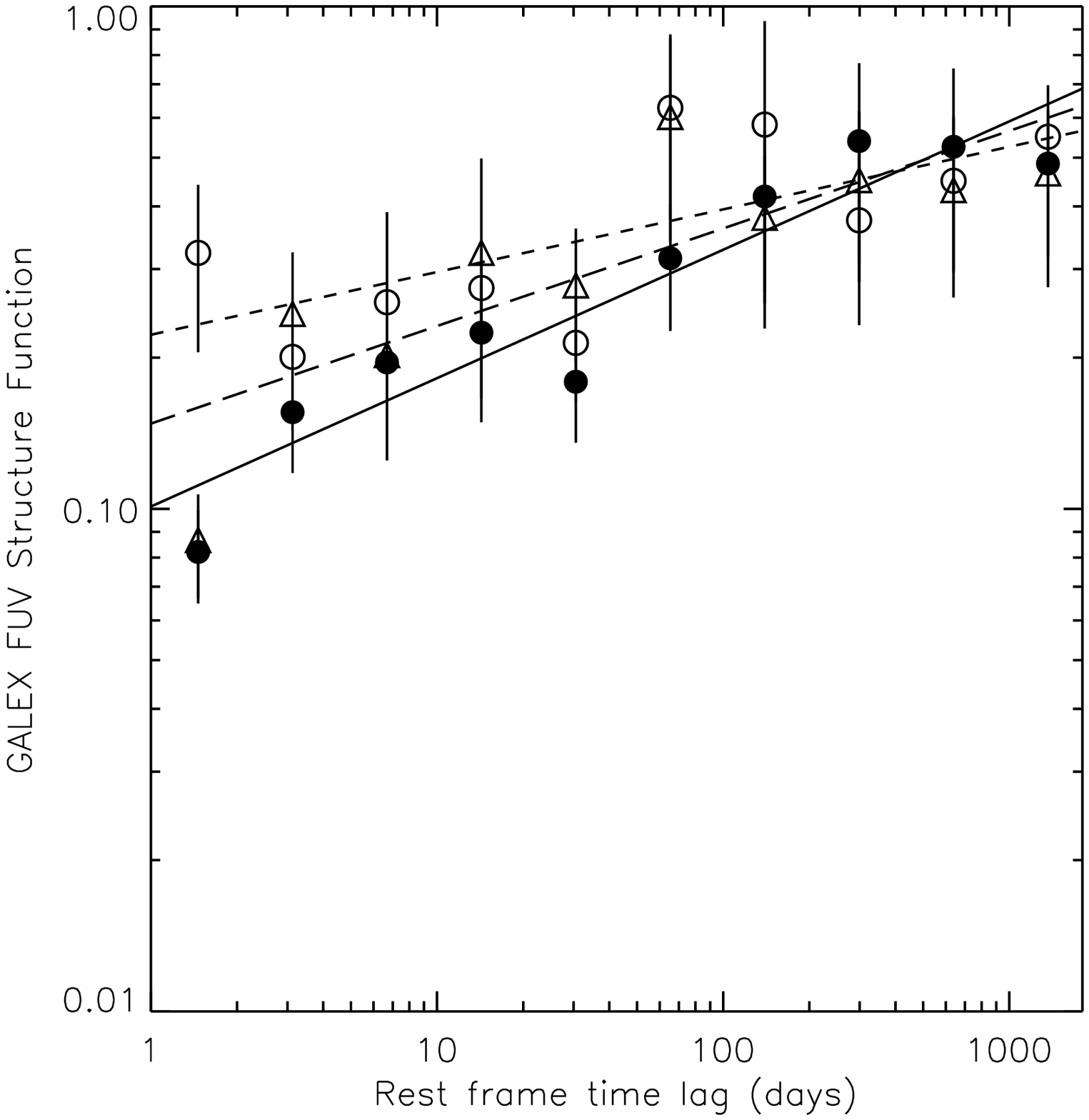}}
\caption{Log-Log plots of the FUV structure function (in 3 redshift bins) with respect to the rest frame time lags for the variable quasars. The data are shown for the same 10 bins of time lags as Figure 7, with the data averages plotted for redshift bins of size 0 $<$ $\it z$  $<$ 0.47 (open circles),  0.47 $<$ $\it z$ $<$ 1.33 (filled circles) and $\it z$ $>$ 1.33 (open triangles) . The vertical lines represent error bars for the binned data points.  Linear best-fits are also shown for each of the redshift bin data sets; long dashed line (open circle data), full line (filled circle data) and short dashed line (open triangle data). None of these linear fits match the FUV data for time-lags $>$ 300 days.}
\label{Figure 9}
\end{figure}

\begin{figure}
\center
{\includegraphics[height=8cm]{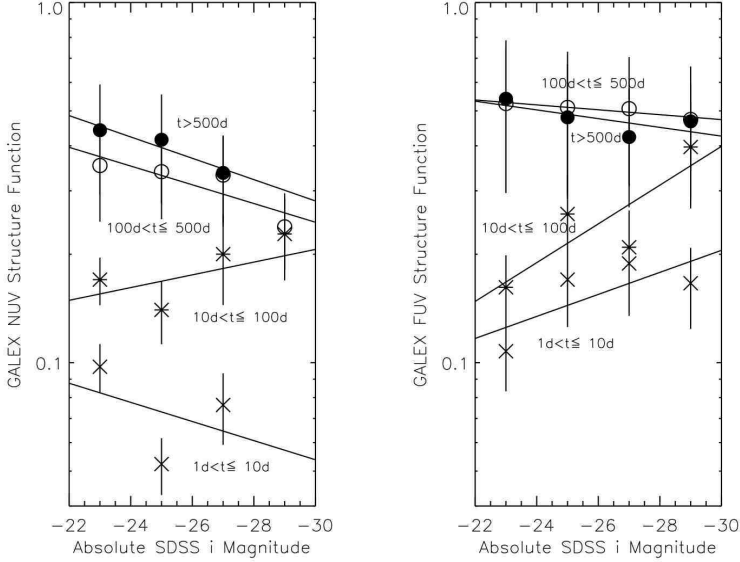}}
\caption{Plots of the NUV (left) and FUV (right) structure functions as a function of quasar luminosity (absolute $\it i$ magnitude) for our sample of UV variable objects. The data are plotted in bins of 2 mag width and are shown for four rest frame time-lags (t) of 1d $<$ t $\le$ 10d (crosses), 10d $<$ t $\le$ 100d (asterisks), 100d $<$ t $\le$ 500d (open circles) and $>$ 500d (filled circles) for both the NUV and FUV observations. Best-fit lines to the data points are also shown, revealing that NUV and FUV variability amplitudes are greater at longer time-lags than for shorter time-lags for all values of QSO luminosity}
\label{Figure 10}
\end{figure}

\begin{figure}
\center
{\includegraphics[height=8cm]{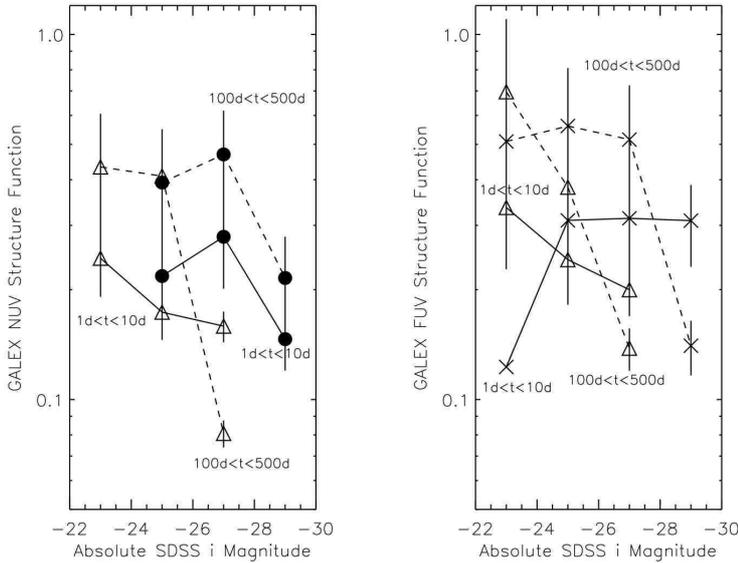}}
\caption{Log-Log plots of the NUV (left) and FUV (right) structure functions as a function of quasar luminosity (absolute $\it i$ magnitude) for our sample of UV variable objects for rest frame time-lags of 1 d $<$ 10d and 100d $<$ 500d for  QSOs with redshifts of 0.5 $<$ $\it z$ $<$ 0.9 (open triangles) and $\it z$ $>$ 2.03 (filled circles) for the NUV data, and redshifts of 0.5 $<$ $\it z$ $<$ 0.9 (open triangles) and 1.4 $<$ $\it z$ $<$ 2.03 (asterisks). The data points have been connected by a continuous line for the 1 d $<$ 10d data, and by a dashed line for the 100d $<$ 500d data for ease of viewing.}
\label{Figure 11}
\end{figure}

\begin{figure}
\center
{\includegraphics[height=10cm]{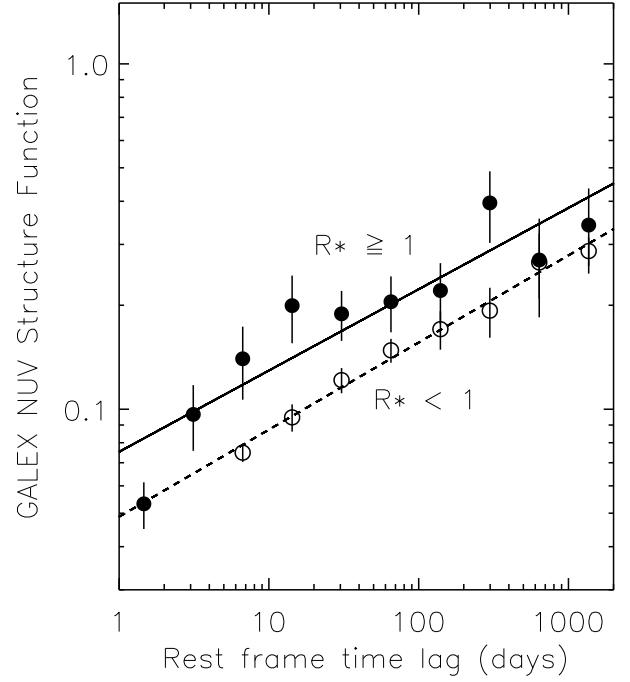}}
\caption{Log-Log plot of the NUV structure function as a function of rest frame time-lag for radio bright UVQs  (R$^{*}$ $>$ 1, filled circles) and radio quiet (R$^{*}$
$<$ 1, open circles) UVQs,
as detected by the VLA 20cm FIRST survey. Best fit straight lines are shown for each of the two data sets.
The radio bright NUV data are best fit with a line of slope 0.25 (intercept 0.05) and the radio faint sources with a similar slope of 0.24 (intercept 0.08).}
\label{Figure 12}
\end{figure}

\begin{figure}
\center
{\includegraphics[height=10cm]{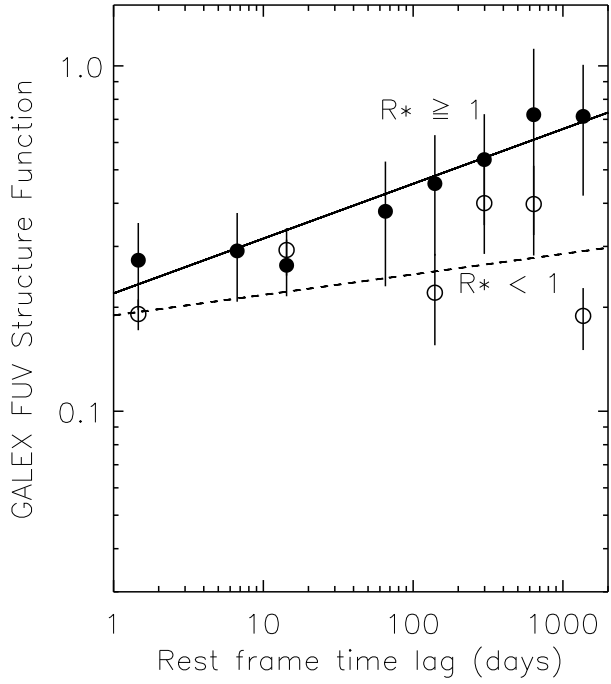}}
\caption{Log-Log plot of the FUV structure function as a function of rest frame time-lag for radio bright UVQs  (R$^{*}$ $>$ 1, filled circles) 
 and radio quiet UVQs (R$^{*}$ $<$ 1, open circles)
as detected by the VLA 20cm FIRST survey. Best fit straight lines are shown for each of the two data sets. The radio bright FUV data are best fit with
 a line of slope 0.16 (intercept 0.22) and the radio faint sources with a slope of 0.059 (intercept 0.19).}
\label{Figure 13}
\end{figure}

\begin{figure}
\center
{\includegraphics[height=8cm]{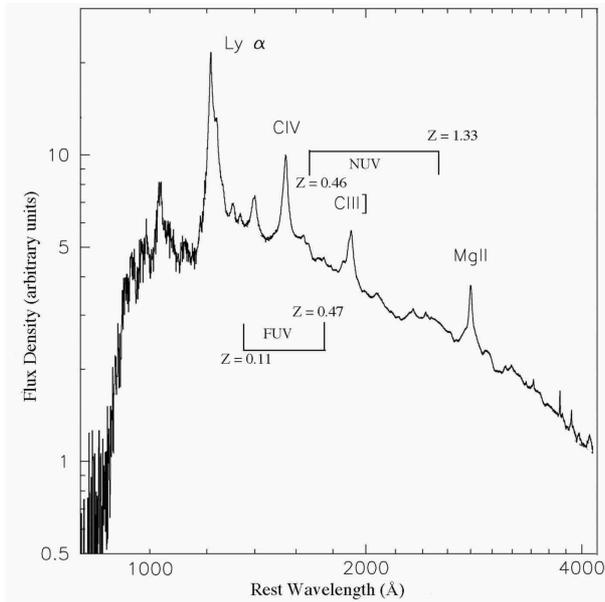}}
\caption{UV-visible spectrum of a zero redshift quasar showing the most prominent emission lines. Also shown are the $\it GALEX$ FUV and NUV passbands together with the redshift (z) values at which Ly-$\alpha$ emission enters and leaves each of the passbands (adapted from Vanden Berk et al. 2001) }
\label{Figure 14}
\end{figure}

\end{document}